\documentclass[12pt,preprint]{aastex}
\usepackage{gensymb}

\shorttitle{C3H2 detection in a disk}
\shortauthors{}

\begin{document}

\title{First detection of $c$-C$_3$H$_2$ in a circumstellar disk}

\author{Chunhua Qi}

\affil{Harvard-Smithsonian Center for Astrophysics, 60 Garden Street, Cambridge, MA 02138, USA} 

\and

\author{Karin I. \"Oberg}

\affil{Departments of Chemistry and Astronomy, University of Virginia, Charlottesville, VA 22904, USA}

\and

\author{David J. Wilner, Katherine A. Rosenfeld}

\affil{Harvard-Smithsonian Center for Astrophysics, 60 Garden Street, Cambridge, MA 02138, USA}

\begin{abstract}

We report the first detection of $c$-C$_3$H$_2$ in a circumstellar disk. 
The $c$-C$_3$H$_2$ J=$6-5$ line (217.882 GHz) is detected and imaged 
through Atacama Large Millimeter Array (ALMA) Science Verification
observations toward the disk around the Herbig Ae star HD~163296 at 
0.8$''$ resolution. The emission is consistent with that arising from
a Keplerian rotating disk. Two additional $c$-C$_3$H$_2$ transitions are 
also tentatively detected, bolstering the identification of this
species, but with insufficient signal-to-noise ratio to constrain the
spatial distribution.  Using a previously developed model for the
physical structure of this disk, we fit a radial power-law
distribution model to the $c$-C$_3$H$_2$ $6-5$ emission and find that
$c$-C$_3$H$_2$ is present in a ring structure from an inner radius of
about 30 AU to an outer radius of about 165 AU.  The column density is
estimated to be 10$^{12}$--10$^{13}$ cm$^{-2}$. The clear detection and
intriguing ring structure  suggest that $c$-C$_3$H$_2$ has 
the potential to become a useful probe of radiation penetration in disks. 

\end{abstract}


\section{Introduction}

Pre-main sequence stars are commonly surrounded by disks, which
serve the dual purpose of funneling accretion onto the central star
and transporting away angular momentum. These disks are also the
formation sites of planets and connect protostars and planets
physically and chemically. The main component of gas-rich
circumstellar disks, cold H$_2$, cannot be observed directly because
of a lack of transitions at suitable energy levels. Instead,
constraints on disk mass, density and temperature structures depend on
observations of trace species, most commonly dust and CO gas
\citep[e.g.][]{Qi06,Pietu07,Andrews09}. To access other disk
properties, such as ionization levels, X-ray and UV radiation fields,
turbulent mixing, chemical age and bulk composition requires the
development of additional molecular probes whose formation and 
destruction depend critically on one or several of these properties.  

The past decade has witnessed significant progress on disk chemistry
theory, resulting in large numbers of predictions on how molecular
abundance structures should depend on different disk properties
\citep[e.g.][]{Aikawa06,Willacy07,Semenov11,Fogel11,Walsh12}. 
A lack of observational sensitivity has however limited detections to
the simplest molecules in the millimeter/submillimeter range which
characterizes the bulk of the gas disk, 
including CO, HCO$^+$, DCO$^+$, CN, HCN, DCN, C$_2$H, H$_2$CO, CS, 
SO, CH$_2$, N$_2$H$^+$ and HC$_3$N
\citep{Dutrey97,Aikawa03,Thi04,Semenov05,Dutrey07,Qi08,Fuente10,
Henning10,Oberg10c,Oberg11a}.
This list is expected to expand with the arrival of ALMA
and its unprecedented sensitivity and spatial resolution, which will
enable the development of a more diverse set of molecular probes.
In this {\em Letter}, we use ALMA Science Verification observations of
HD 163296 to report the first detection of cyclopropenylidene, 
$c$-C$_3$H$_2$, in a circumstellar disk. At a distance of $\sim$122
pc (\citealp{perryman_l97}), this Herbig Ae star 
(stellar mass 2.3 M$_{\odot}$; spectral type
A1; age $\sim$4 Myr) harbors a large disk (a scattered light pattern
extending out to a radius of $\sim$500 AU, \citealp{grady_d00}) with strong
millimeter continuum and molecular line emission
(\citealp{mannings_s97,Thi04,natta_t04,isella_t07,Qi11}). The large
disk size and strong molecular emission make this system an ideal
target to search for new molecules.

The $c$-C$_3$H$_2$ molecule is a small cyclic hydrocarbon, which was 
first detected in space by \citet{Thaddeus85} towards Sgr B2. Since 
then, $c$-C$_3$H$_2$ has been detected in a wide range of astrophysical
environments, including diffuse and dense clouds, protostars,
planetary nebulae and extragalactic sources
\citep[e.g.,][]{Matthews85,Cox88,Cox89,Madden89,Benson98,Menten99,Fosse01,Sakai08,Tenenbaum09,Liszt12},
and is one of the most abundant molecules with 3 carbon atoms in the  
interstellar medium \citep{Teyssier04}.  In the context of disks,
$c$-C$_3$H$_2$ has some favorable characteristics that may make it a
powerful molecular probe, especially when observed in combination with
isotopologues and chemically related molecules:
\begin{enumerate}
\item $c$-C$_3$H$_2$ has two equivalent H nuclei, which couple to
  generate ortho (nuclear spin of 1) and para (nuclear spin of 0)
  states. Deviations from the statistical weights of 3 to 1 may be
  used to constrain the formation conditions of the molecule
  \citep{Park06}.  
\item $c$-C$_3$H$_2$ is also one of two isomers; the other is the
  linear molecule $l$-C$_3$H$_2$. The isomer ratio can be used to
  assess the fractional ionization in clouds \citep{Fosse01} and may
  provide similar constraints on disks.  
\item The formation and destruction chemistry of $c$-C$_3$H$_2$ (and
  carbon chains)  are sensitive to high energy radiation. The
  vertical and radial abundance structure should provide strong constraints 
  on the penetration depth of X-rays and UV photons \citep{Semenov11}. 
\item Within the family of hydrocarbons, the sensitivity to turbulent
  mixing increases dramatically from C$_2$ to C$_3$H$_2$, and from
  C$_3$H$_2$ to larger hydrocarbons. Observed ratios and limits have
  the potential to constrain transport in disks
  \citep{Semenov11}. 
\end{enumerate}

Finally, the chemistry that produces $c$-C$_3$H$_2$ is important to
constrain in its own right, since $c$-C$_3$H$_2$ shares precursors
with other, more complex carbon chains, which may be an important
source of carbon during planet formation. In light of these
possibilities, we aim to provide first constraints on the abundance,
distribution and $o/p$ ratio of $c$-C$_3$H$_2$ in the HD~163296 disk. 

\section{Observations\label{sec:obs}}

HD~163296 (RA: 17$^{\rm h}$56$^{\rm m}$21\fs287, DEC: 
$-$21\degr57$'$22\farcs39; J2000.0) was observed on 2012 June 9, June
23 and July 7 as part of 
the ALMA Science Verification (SV) program in band 6. At the time of
observations, 25 12-m antennae were used (2 7-m ACA antennae were 
present in the dataset but flagged) in the array with a resulting
synthesized beam size of 0.9$\times$0.7$\arcsec$ (PA=84\degree). 
The source was observed with a total integration time of 84 minutes. 
The ALMA correlator was configured to simultaneously observe four 
spectral windows (two in each sideband). One of those windows (SpwID
\#1) covered both $^{13}$CO and C$^{18}$O $2-1$ lines (220.399 and
219.560 GHz, respectively).  
The data were calibrated in the CASA software package (v3.4),
following the detailed calibration and imaging scripts provided by the
ALMA science verification team. Since those scripts, along with
fully calibrated measurement sets, are publicly available
online\footnote{https://almascience.nrao.edu/almadata/sciver/HD163296Band6},
we do not repeat the details here. 
The 218 GHz (1.37mm) continuum was generated
from the line-free channels in the two spectral windows
(SpWID \#0 and \#1) of the lower sideband, 
and the flux is determined to be  608.5$\pm$2.5 mJy,
which agrees with the SMA observations \citep{Qi11}.    
Excellent agreement is also seen between the SMA and ALMA
spectra for the $^{13}$CO and C$^{18}$O $2-1$ lines with fluxes within 10\%
 \citep{Qi11}.

All of the $c$-C$_3$H$_2$ lines
are continuum-subtracted and retrieved from SpwID \#0 with a channel
width of 0.488 MHz (0.67 km s$^{-1}$) covering the total 1.875 GHz
bandwidth from 216.167 to 218.042 GHz. We have binned the lines in
channels of 1 km s$^{-1}$ to achieve higher signal-to-noise ratios and
still keep the key kinematic features of the disk.  The resulting line
rms noise level is estimated to be 2.2 mJy beam$^{-1}$ per 1 km
s$^{-1}$ channel in line free channels.  

\section{Results and Analysis}\label{sec:res}

\subsection{$c$-C$_3$H$_2$ detection}

\begin{figure}
\epsscale{1.0}
\plotone{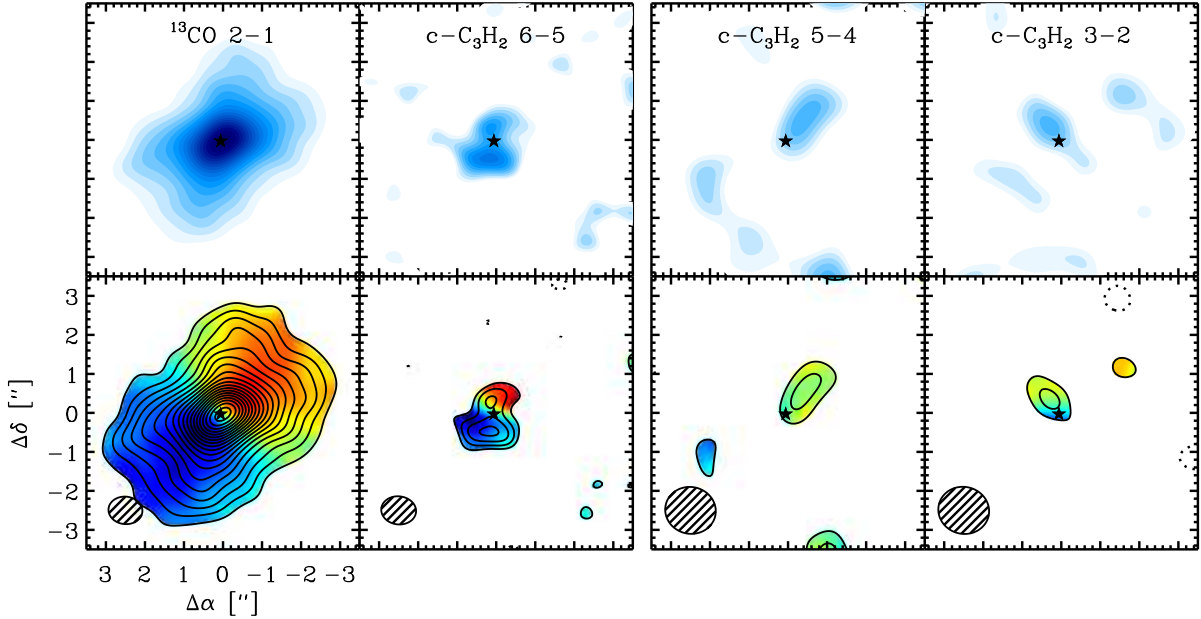}
\caption{The integrated intensity maps summed between 0 and 11 km
  s$^{-1}$) and intensity-weighted mean 
  velocity fields of $^{13}$CO $2-1$ and $c$-C$_3$H$_2$ $6-5$ lines
  (left panel), $c$-C$_3$H$_2$ $5-4$ and $3-2$ lines (right panel)
  toward HD~163296. The resolved velocity field of the
  $c$-C$_3$H$_2$ $6-5$ line agrees with the CO
  kinematics. In the $c$-C$_3$H$_2$ maps, the first contour marks
  3$\sigma$ followed by 1$\sigma$ contour increases. The rms varies
  between 6 and 9 mJy km s$^{-1}$ per beam. Synthesized beams are
  presented in the lower left corners. The star symbol indicates the
  continuum (stellar) position. The axes are offsets from the pointing
  center in arcseconds.\label{fig:mom}} 
\end{figure}

Figure \ref{fig:mom} shows that the $c$-C$_3$H$_2$ $6-5$ line is
clearly detected towards HD~163296 and that the position angle of its
resolved velocity field is consistent with that of $^{13}$CO from the
same data set, which shows a clear pattern of Keplerian rotation. 
Interestingly, the $c$-C$_3$H$_2$  $6-5$ moment map
displays a double-peaked morphology, indicative of an inner hole. 
The $c$-C$_3$H$_2$ $3-2$ and $5-4$ lines are only tentatively
detected, insufficient to resolve the velocity field or any 
spatial structure. To improve the sensitivity in imaging these lines, 
we applied a Gaussian taper to the visibility data that corresponds 
to 1\arcsec ~FWHM in the image domain, 
which resulted in a beam of 1.3$\times$1.2$\arcsec$ (PA=78\degree).

\begin{figure}
\epsscale{0.8}
\plotone{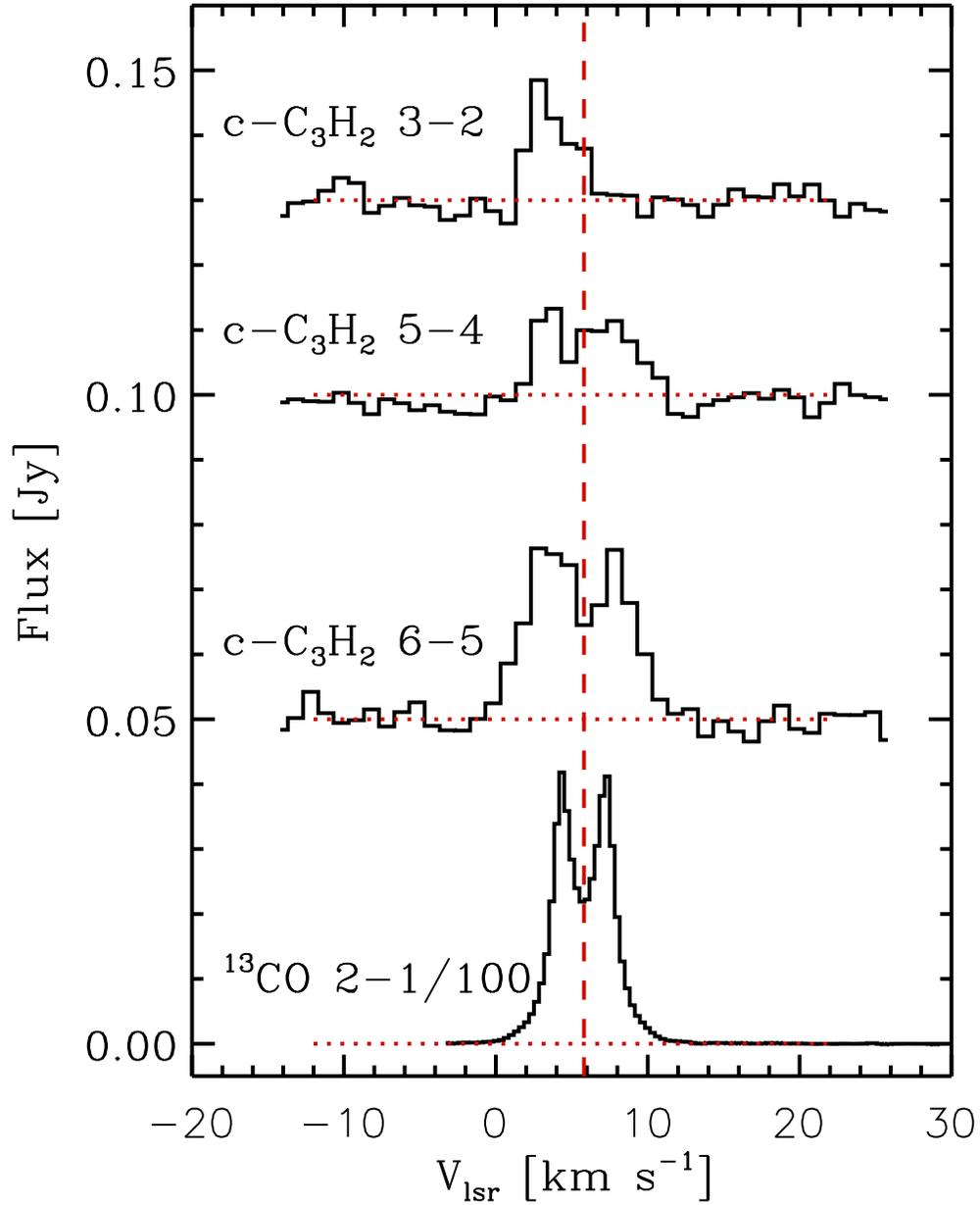}
\caption{Extracted spectra of the three $c$-C$_3$H$_2$ transitions
  toward HD~163296,
  plotted together with $^{13}$CO $2-1$. The dashed line marks V$_{\rm
  LSR}$ toward HD~163296.
\label{fig:spec}}
\end{figure}

Because the line emission changes position with channels, using
a single mask for all channels to extract spectra will lead to a
degradation of the signal. Instead, we create 
separate masks to cover the $c$-C$_3$H$_2$ $6-5$ line emission at each
channel and use these masks to extract the line spectra. The same masks are
also used for other $c$-C$_3$H$_2$ lines. 
Figure \ref{fig:spec}
shows the resulting spectra of all three $c$-C$_3$H$_2$ $3-2$ lines
together with $^{13}$CO $2-1$ line (with its own masks
covering the $^{13}$CO $2-1$ emission at each channel). 
The $6-5$ line is clearly detected with the expected line shape and 
V$_{\rm lsr}$ consistent with $^{13}$CO. 
The $5-4$ line presents a consistent spectral profile, at lower
signal-to-noise ratio. In contrast, the 
$3-2$ line is both narrower and clearly blue-shifted, suggestive of a
different physical origin, perhaps in a disk wind. However, both the 
$3-2$ and $5-4$ lines are observed at the edge of the spectral window, 
so this apparent asymmetry should be confirmed by independent observations
before any strong conclusions are drawn. 

The integrated fluxes are reported in Table \ref{tab:line}. 
These are just below the upper limits reported in two previous failed 
searches of $c$-C$_3$H$_2$ in disks  \citep{Fuente10,Oberg10c}.

\subsection{Ring structure model}

We explore the distributions of $c$-C$_3$H$_2$ in HD~163296 based on a
previously developed accretion disk model with well-defined
temperature and density structures \citep{Qi11}.
To simulate and compare with the data, we assume the disk material
orbits the central star in Keplerian motion, and fix the disk geometric
and kinematic parameters that affect the observed spatio-kinematic
behavior of the disk. The details of the stellar and accretion
properties and the disk geometric and kinematic parameters are
summarized in Table 3 of \citet{Qi11}. We have developed a physically
self-consistent accretion disk model with an exponentially tapered
edge that matches the spectral energy distribution and spatially
resolved millimeter dust continuum emission. The disk temperature and
density structures are further constrained by multiple CO and CO
isotopologue line observations. Such analysis provides
the essential framework for constraining the distribution of molecular
abundances in protoplanetary disks. 

Within this model framework, we assume that $c$-C$_3$H$_2$ is present 
with a constant abundance in a layer with boundaries toward the 
midplane and toward the surface of the disk \citep{Qi08,Oberg12}.
This assumption is motivated by 
chemical models \citep[e.g.][]{Aikawa06} that predict a three-layered 
structure where most molecules are photodissociated in the surface layer, 
frozen out in the midplane, and have an abundance that peaks at 
intermediate disk heights. The surface ($\sigma_s$) and midplane 
($\sigma_m$) boundaries are presented in terms of 
$\Sigma_{21}=\Sigma_H/(1.59\times10^{21} cm^{-2})$, where $\Sigma_H$
is the hydrogen column density measured from the disk surface in the
adopted physical model. This simple model approach approximates the
vertical location where $c$-C$_3$H$_2$ is most abundant.
Due to the limited signal-to-noise of the multiple $c$-C$_3$H$_2$ line 
detections, we cannot constrain the 
location of this vertical layer based on the $c$-C$_3$H$_2$ data.  
We therefore fix the vertical surface boundary $\sigma_s$ as 0.32
($\log_{10}(\sigma_s)=-0.5$) and the midplane boundary $\sigma_m$ as 3.2
($\log_{10}(\sigma_m)=0.5$), between which lies the expected location of  most
molecules with peak abundances in the warm molecular layer
\citep{Aikawa06}.

We model the radial column density distribution of $c$-C$_3$H$_2$ as a
power law N$_{100}\times(r/100)^p$, where N$_{100}$ is the column density at
100 AU in cm$^{-2}$, $r$ is the distance from the star in AU, and $p$ 
is the power-law index. Since the emission shows a clear indication of 
a central hole, we fit for both an inner radius R$_{in}$ and outer radius
R$_{out}$ together with the power-law parameters 
(N$_{100}$ and $p$). Using the structure model and the fixed vertical
distributions, we 
compute a grid of synthetic $c$-C$_3$H$_2$ J=$6-5$ visibility datasets over a
range of R$_{out}$, R$_{in}$, $p$ and N$_{100}$ values and compare
with the observations. The best-fit model is obtained by minimizing
$\chi^2$, the weighted  
difference between the real and imaginary part of the complex visibility 
measured in the ($u,v$)-plane sampled by the ALMA observations.
We use the two-dimensional Monte Carlo model RATRAN \citep{Hogerheijde00} 
to calculate the radiative transfer and molecular excitation. 
The collisional rates are taken from \citet{Chandra00} and the
molecular data file is retrieved from the Leiden Atomic and Molecular 
Database \citep{Schoier05}.

\begin{figure}
\epsscale{0.8}
\plotone{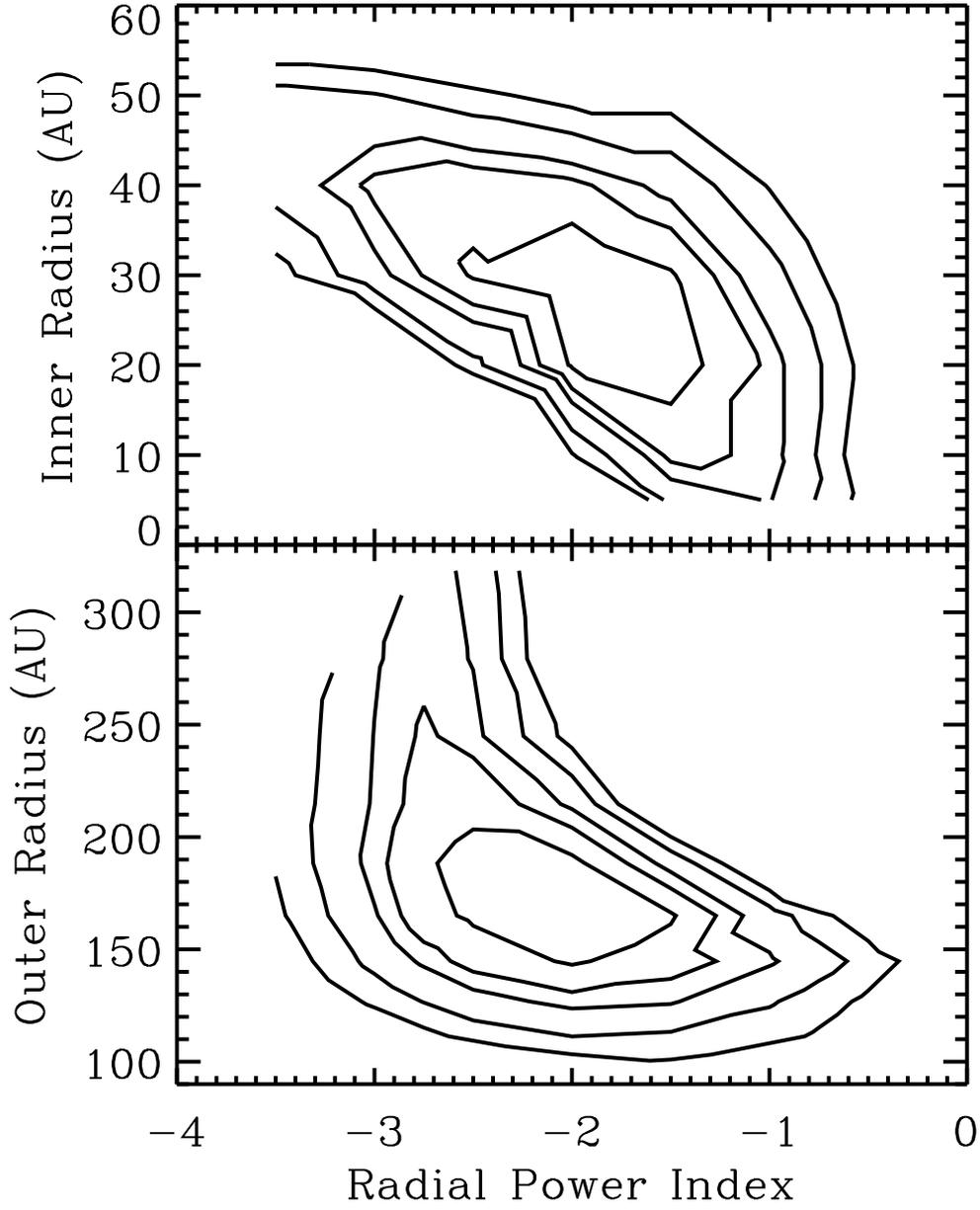}
\caption{Iso-$\chi^2$ surfaces of $R_{\rm out}$ and $R_{\rm in}$
  versus $p$. Contours correspond to the 1--5 $\sigma$ errors. 
\label{fig:chi2}}
\end{figure}

\begin{figure}
\epsscale{1.0}
\plotone{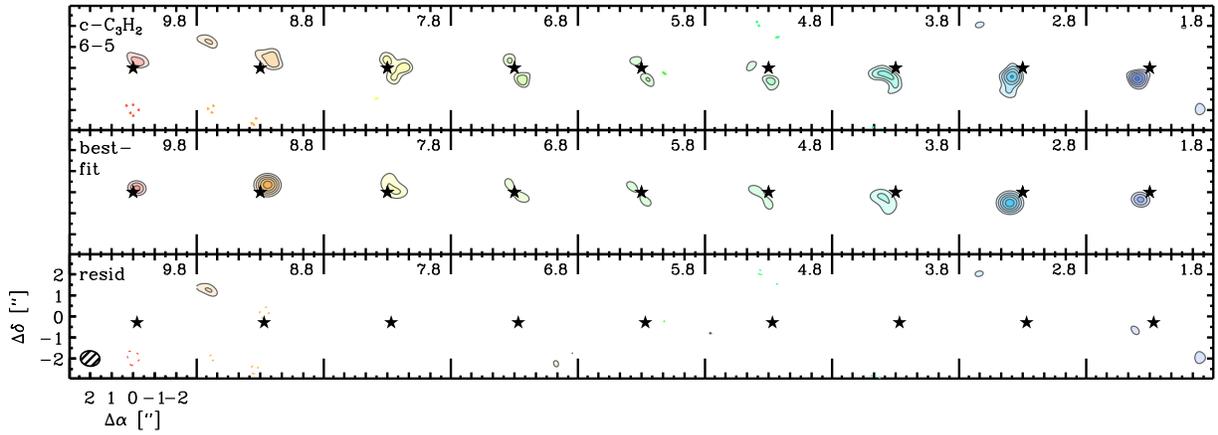}
\caption{The top panel shows the velocity channel map of the
  $c$-C$_3$H$_2$ $6-5$ emission toward HD~163296 (velocities binned in
  1 km s$^{-1}$). Contours are 0.0022 Jy Beam$^{-1}$ (1$\sigma$) $\times
  [3,4,5,6,7,8,9]$. The middle and bottom panels show the best-fit model and
  the difference between the best-fit model and data on the
  same contour scale. The star symbol indicates the continuum (stellar)
  position. The axes are offsets from the pointing center in arcseconds.
\label{fig:chmap}}
\end{figure}

Figure~\ref{fig:chi2} shows the $\chi^2$ surfaces for the R$_{in}$ and
$R_{out}$ versus the power law index $p$. We find the $p$ is constrained
between -2.5 and -1.5, R$_{in}$ 15--40 AU and R$_{out}$ 150--200 AU 
(within 1$\sigma$) with best-fit values at -2, 30 and 165 AU, respectively
(reduced $\chi^2=1.07$). 
By fixing the above parameters at their best-fit values, 
N($c$-C$_3$H$_2$) is determined to be 
2.2$\pm$0.2 $\times$10$^{12}$ cm$^{-2}$ at 100 AU. 
Figure~\ref{fig:chmap} presents comparisons between the observed
channel maps and the best-fit model. 
The residual image doesn't show any significant emission, which 
indicates that the $c$-C$_3$H$_2$ emission is consistent
with that arising from a Keplerian rotating disk.
The $6-5$ line is actually blended
with the $6_{ 1, 6}- 5_{ 0, 5}$ (ortho) and $6_{0, 6}- 5_{1, 5}$ (para) lines at the
same frequency. Since the Einstein $A$-coefficients for both ortho and para
lines are the same, 5.393$\times$10$^{-4}$ s$^{-1}$
\citep{Schoier05}, we fit the data with the ortho line only and 
the resulting column density can be accounted as the sum of both ortho and
para forms of the molecule. 

For the weaker $5-4$ ortho line, the signal-to-noise ratio is too low to 
provide any constraints on the $c$-C$_3$H$_2$ the distribution. By assuming 
the $c$-C$_3$H$_2$ distribution derived from the $6-5$ line, however, 
we can use the unresolved line flux to constrain the $o/p$ ratio. 
The resulting best-fit 
N(C$_3$H$_2$-($o$))=1.6$\pm$0.3 $\times$10$^{12}$ and
N(C$_3$H$_2$-($o+p$)=2.2$\pm$0.2 $\times$10$^{12}$ cm$^{-2}$ at 100 AU 
(as shown above). Hence
 $o/p$=2.7$\pm$1.7, which is close to the statistical value of 3, but
the large uncertainty implies that better data are
required to use this ratio as a probe of formation conditions.  

\section{Discussion\label{sec:disc}}

We have clearly detected $c$-C$_3$H$_2$ in the disk around HD~163296. 
This is one of the two largest molecules detected in disks so far,
the other being HC$_3$N, also a carbon-chain molecule, detected recently 
using deep single-dish observations \citep{Chapillon12}. 

Using the spatially and spectrally resolved line emission, we have 
constrained the radial distribution of $c$-C$_3$H$_2$ in the HD~163296 
disk as a ring structure ranging from $\sim$30 to $\sim$165 AU. 
The outer edge of  $<$250 AU (2$\sigma$
limit) is much smaller than the gas disk size which extends to 500 AU 
\citep{Qi11}, which suggests that $c$-C$_3$H$_2$ formation is slow or 
that $c$-C$_3$H$_2$ destruction is fast in the outer disk. 
In terms of formation, the best-fit outer radius coincides with the 
previously determined location of the CO midplane snow line, and it is 
possible that CO freeze-out limits the carbon available in the gas phase 
to form hydrocarbons in general and $c$-C$_3$H$_2$ in particular. A second characteristic of the outer disk is that 
it is rather tenuous because
of a rapidly decreasing column density outside of 200~AU
\citep{Qi11}. This may result in efficient penetration of radiation
and thus destruction of the easily dissociated
$c$-C$_3$H$_2$. Explicit model predictions are required to resolve which effect drives the disappearance of
$c$-C$_3$H$_2$ in the outer disk. 

No existing disk chemistry model in the literature contains predictions
for the radial distribution of $c$-C$_3$H$_2$.  Based on strong correlations 
found between $c$-C$_3$H$_2$ and CCH in several astrophysical
environments \citep{Lucas00,Pety05,Gerin11}, predictions on CCH in
disks can be used to shed further light on the origins of $c$-C$_3$H$_2$.
\citet{Aikawa01} predicts a CCH central cavity in the absence of
X-ray radiation; when X-rays are present, and HD 163296 is a
known X-ray emitter \citep{Swartz05,Guenther09}, the CCH column
is instead centrally peaked. Despite the presence of X-rays, the
central cavity could be due to a UV dominated radiation field.
This may be tested by observing $c$-C$_3$H$_2$ towards T Tauri
stars with weak UV fields and strong X-rays, and towards additional
Herbig Ae stars with weak X-rays.

The inner cavity could also be a product of the main formation pathway
of $c$-C$_3$H$_2$. Carbon chains can form through at least
three different pathways: (1) ion-neutral reactions at low temperatures,
(2) photo-erosion of larger carbonaceous compounds, and (3) neutral-neutral
reactions at luke-warm temperatures following CH$_4$ ice evaporation
at $\sim$30~K  \citep{Herbst84,Teyssier04,Sakai08,Gerin11}. Detailed
modeling of $c$-C$_3$H$_2$ radial distributions in each of these
formation scenarios are clearly needed to use $c$-C$_3$H$_2$ rings as
tracers of the radiation field and other disk characteristics. 

In addition to more detailed modeling, additional observations are
needed to constrain whether different hydrocarbons, especially CCH and
$c$-C$_3$H$_2$, are correlated in disk environments, and to constrain 
the vertical distribution of these molecules. The latter requires multiple 
lines with a range of excitation energies. With the combined advance in
theory and observations, $c$-C$_3$H$_2$ may become one of the more
useful probes of penetration of the radiation fields in disks. 

{\it Facility:} \facility{ALMA}

\acknowledgments

\noindent  

This paper makes use of the following ALMA data:
ADS/JAO.ALMA\#2011.0.00010.SV. ALMA is a partnership of ESO
(representing its member states), NSF (USA) and NINS (Japan), together
with NRC (Canada) and NSC and ASIAA (Taiwan), in cooperation with the
Republic of Chile. The Joint ALMA Observatory is operated by ESO,
AUI/NRAO and NAOJ. \\
We would like to thank an anonymous referee for thoughtful suggestions
for the paper. We also acknowledge NASA Origins of Solar Systems grant
No. NNX11AK63.

\bibliographystyle{aa}

\begin{thebibliography}{49}
\expandafter\ifx\csname natexlab\endcsname\relax\def\natexlab#1{#1}\fi

\bibitem[{{Aikawa} \& {Herbst}(2001)}]{Aikawa01}
{Aikawa}, Y. \& {Herbst}, E. 2001, \aap, 371, 1107

\bibitem[{{Aikawa} {et~al.}(2003){Aikawa}, {Momose}, {Thi}, {van Zadelhoff},
  {Qi}, {Blake}, \& {van Dishoeck}}]{Aikawa03}
{Aikawa}, Y., {Momose}, M., {Thi}, W., {et~al.} 2003, \pasj, 55, 11

\bibitem[{{Aikawa} \& {Nomura}(2006)}]{Aikawa06}
{Aikawa}, Y. \& {Nomura}, H. 2006, \apj, 642, 1152

\bibitem[{{Andrews} {et~al.}(2009){Andrews}, {Wilner}, {Hughes}, {Qi}, \&
  {Dullemond}}]{Andrews09}
{Andrews}, S.~M., {Wilner}, D.~J., {Hughes}, A.~M., {Qi}, C., \& {Dullemond},
  C.~P. 2009, \apj, 700, 1502

\bibitem[{{Benson} {et~al.}(1998){Benson}, {Caselli}, \& {Myers}}]{Benson98}
{Benson}, P.~J., {Caselli}, P., \& {Myers}, P.~C. 1998, \apj, 506, 743

\bibitem[Chandra 
\& Kegel(2000)]{Chandra00} Chandra, S., \& Kegel, W.~H.\ 2000, \aaps, 142, 113 

\bibitem[{{Chapillon} {et~al.}(2012){Chapillon}, {Dutrey}, {Guilloteau},
  {Pi{\'e}tu}, {Wakelam}, {Hersant}, {Gueth}, {Henning}, {Launhardt},
  {Schreyer}, \& {Semenov}}]{Chapillon12}
{Chapillon}, E., {Dutrey}, A., {Guilloteau}, S., {et~al.} 2012, \apj, 756, 58

\bibitem[Cox et 
al.(1988)]{Cox88} Cox, P., Guesten, R., \& Henkel, C.\ 1988, \aap, 206, 108 

\bibitem[Cox et 
al.(1989)]{Cox89} Cox, P., Walmsley, C.~M., \& Guesten, R.\ 1989, \aap, 209, 382 

\bibitem[{{Dutrey} {et~al.}(1997){Dutrey}, {Guilloteau}, \&
  {Guelin}}]{Dutrey97}
{Dutrey}, A., {Guilloteau}, S., \& {Guelin}, M. 1997, \aap, 317, L55

\bibitem[{{Dutrey} {et~al.}(2007){Dutrey}, {Henning}, {Guilloteau}, {Semenov},
  {Pi{\'e}tu}, {Schreyer}, {Bacmann}, {Launhardt}, {Pety}, \&
  {Gueth}}]{Dutrey07}
{Dutrey}, A., {Henning}, T., {Guilloteau}, S., {et~al.} 2007, \aap, 464, 615

\bibitem[{{Fogel} {et~al.}(2011){Fogel}, {Bethell}, {Bergin}, {Calvet}, \&
  {Semenov}}]{Fogel11}
{Fogel}, J.~K.~J., {Bethell}, T.~J., {Bergin}, E.~A., {Calvet}, N., \&
  {Semenov}, D. 2011, \apj, 726, 29

\bibitem[{{Foss{\'e}} {et~al.}(2001){Foss{\'e}}, {Cernicharo}, {Gerin}, \&
  {Cox}}]{Fosse01}
{Foss{\'e}}, D., {Cernicharo}, J., {Gerin}, M., \& {Cox}, P. 2001, \apj, 552,
  168

\bibitem[{{Fuente} {et~al.}(2010){Fuente}, {Cernicharo}, {Ag{\'u}ndez},
  {Bern{\'e}}, {Goicoechea}, {Alonso-Albi}, \& {Marcelino}}]{Fuente10}
{Fuente}, A., {Cernicharo}, J., {Ag{\'u}ndez}, M., {et~al.} 2010, \aap, 524,
  A19

\bibitem[{{Gerin} {et~al.}(2011){Gerin}, {Ka{\'z}mierczak}, {Jastrzebska},
  {Falgarone}, {Hily-Blant}, {Godard}, \& {de Luca}}]{Gerin11}
{Gerin}, M., {Ka{\'z}mierczak}, M., {Jastrzebska}, M., {et~al.} 2011, \aap,
  525, A116

\bibitem[{{Grady} {et~al.}(2000){Grady}, {Devine}, {Woodgate}, {Kimble},
  {Bruhweiler}, {Boggess}, {Linsky}, {Plait}, {Clampin}, \&
  {Kalas}}]{grady_d00}
{Grady}, C.~A., {et~al.} 2000, \apj, 544, 895

\bibitem[G{\"u}nther 
\& Schmitt(2009)]{Guenther09} G{\"u}nther, H.~M., \& Schmitt, J.~H.~M.~M.\ 2009, \aap, 494, 1041 

\bibitem[{{Henning} {et~al.}(2010){Henning}, {Semenov}, {Guilloteau}, {Dutrey},
  {Hersant}, {Wakelam}, {Chapillon}, {Launhardt}, {Pietu}, \&
  {Schreyer}}]{Henning10}
{Henning}, T., {Semenov}, D., {Guilloteau}, S., {et~al.} 2010, \apj, 714, 1511

\bibitem[{{Herbst} {et~al.}(1984){Herbst}, {Adams}, \& {Smith}}]{Herbst84}
{Herbst}, E., {Adams}, N.~G., \& {Smith}, D. 1984, \apj, 285, 618

\bibitem[{{Hogerheijde} \& {van der Tak}(2000)}]{Hogerheijde00}
{Hogerheijde}, M.~R. \& {van der Tak}, F.~F.~S. 2000, \aap, 362, 697

\bibitem[{{Isella} {et~al.}(2007){Isella}, {Testi}, {Natta}, {Neri}, {Wilner},
  \& {Qi}}]{isella_t07}
{Isella}, A., {Testi}, L., {Natta}, A., {Neri}, R., {Wilner}, D., \& {Qi}, C.
  2007, \aap, 469, 213

\bibitem[{{Liszt} {et~al.}(2012){Liszt}, {Sonnentrucker}, {Cordiner}, \&
  {Gerin}}]{Liszt12}
{Liszt}, H., {Sonnentrucker}, P., {Cordiner}, M., \& {Gerin}, M. 2012, \apjl,
  753, L28

\bibitem[{{Lucas} \& {Liszt}(2000)}]{Lucas00}
{Lucas}, R. \& {Liszt}, H.~S. 2000, \aap, 358, 1069

\bibitem[{{Mannings} \& {Sargent}(1997)}]{mannings_s97}
{Mannings}, V., \& {Sargent}, A.~I. 1997, \apj, 490, 792

\bibitem[{{Matthews} \& {Irvine}(1985)}]{Matthews85}
{Matthews}, H.~E. \& {Irvine}, W.~M. 1985, \apjl, 298, L61

\bibitem[Madden et al.(1989)]{Madden89} Madden, S.~C., Irvine, 
W.~M., Swade, D.~A., Matthews, H.~E., \& Friberg, P.\ 1989, \aj, 97, 1403 

\bibitem[{{Menten} {et~al.}(1999){Menten}, {Carilli}, \& {Reid}}]{Menten99}
{Menten}, K.~M., {Carilli}, C.~L., \& {Reid}, M.~J. 1999, in Astronomical
  Society of the Pacific Conference Series, Vol. 156, Highly Redshifted Radio
  Lines, ed. C.~L. {Carilli}, S.~J.~E. {Radford}, K.~M. {Menten}, \& G.~I.
  {Langston}, 218

\bibitem[{{Natta} {et~al.}(2004){Natta}, {Testi}, {Neri}, {Shepherd}, \&
  {Wilner}}]{natta_t04}
{Natta}, A., {Testi}, L., {Neri}, R., {Shepherd}, D.~S., \& {Wilner}, D.~J.
  2004, \aap, 416, 179


\bibitem[{{{\"O}berg} {et~al.}(2010){{\"O}berg}, {Qi}, {Fogel}, {Bergin},
  {Andrews}, {Espaillat}, {van Kempen}, {Wilner}, \& {Pascucci}}]{Oberg10c}
{{\"O}berg}, K.~I., {Qi}, C., {Fogel}, J.~K.~J., {et~al.} 2010, \apj, 720, 480

\bibitem[{{{\"O}berg} {et~al.}(2011){{\"O}berg}, {Qi}, {Fogel}, {Bergin},
  {Andrews}, {Espaillat}, {Wilner}, {Pascucci}, \& {Kastner}}]{Oberg11a}
{{\"O}berg}, K.~I., {Qi}, C., {Fogel}, J.~K.~J., {et~al.} 2011, \apj, 734, 98

\bibitem[{{{\"O}berg} {et~al.}(2012){{\"O}berg}, {Qi}, {Wilner}, \&
  {Hogerheijde}}]{Oberg12}
{{\"O}berg}, K.~I., {Qi}, C., {Wilner}, D.~J., \& {Hogerheijde}, M.~R. 2012,
  ApJ, 749, 162

\bibitem[{{Perryman} {et~al.}(1997){Perryman}, {Lindegren}, {Kovalevsky},
  {Hoeg}, {Bastian}, {Bernacca}, {Cr{\'e}z{\'e}}, {Donati}, {Grenon}, {van
  Leeuwen}, {van der Marel}, {Mignard}, {Murray}, {Le Poole}, {Schrijver},
  {Turon}, {Arenou}, {Froeschl{\'e}}, \& {Petersen}}]{perryman_l97}
{Perryman}, M.~A.~C., {et~al.} 1997, \aap, 323, L49

\bibitem[{{Park} {et~al.}(2006){Park}, {Wakelam}, \& {Herbst}}]{Park06}
{Park}, I.~H., {Wakelam}, V., \& {Herbst}, E. 2006, \aap, 449, 631

\bibitem[{{Pety} {et~al.}(2005){Pety}, {Teyssier}, {Foss{\'e}}, {Gerin},
  {Roueff}, {Abergel}, {Habart}, \& {Cernicharo}}]{Pety05}
{Pety}, J., {Teyssier}, D., {Foss{\'e}}, D., {et~al.} 2005, \aap, 435, 885

\bibitem[{{Pi{\'e}tu} {et~al.}(2007){Pi{\'e}tu}, {Dutrey}, \&
  {Guilloteau}}]{Pietu07}
{Pi{\'e}tu}, V., {Dutrey}, A., \& {Guilloteau}, S. 2007, \aap, 467, 163

\bibitem[{{Qi} {et~al.}(2011){Qi}, {D'Alessio}, {{\"O}berg}, {Wilner},
  {Hughes}, {Andrews}, \& {Ayala}}]{Qi11}
{Qi}, C., {D'Alessio}, P., {{\"O}berg}, K.~I., {et~al.} 2011, \apj, 740, 84

\bibitem[{{Qi} {et~al.}(2008){Qi}, {Wilner}, {Aikawa}, {Blake}, \&
  {Hogerheijde}}]{Qi08}
{Qi}, C., {Wilner}, D.~J., {Aikawa}, Y., {Blake}, G.~A., \& {Hogerheijde},
  M.~R. 2008, \apj, 681, 1396

\bibitem[{{Qi} {et~al.}(2006){Qi}, {Wilner}, {Calvet}, {Bourke}, {Blake},
  {Hogerheijde}, {Ho}, \& {Bergin}}]{Qi06}
{Qi}, C., {Wilner}, D.~J., {Calvet}, N., {et~al.} 2006, \apjl, 636, L157

\bibitem[{{Sakai} {et~al.}(2008){Sakai}, {Sakai}, \& {Yamamoto}}]{Sakai08}
{Sakai}, N., {Sakai}, T., \& {Yamamoto}, S. 2008, \apss, 313, 153

\bibitem[{{Sch{\"o}ier} {et~al.}(2005){Sch{\"o}ier}, {van der Tak}, {van
  Dishoeck}, \& {Black}}]{Schoier05}
{Sch{\"o}ier}, F.~L., {van der Tak}, F.~F.~S., {van Dishoeck}, E.~F., \&
  {Black}, J.~H. 2005, \aap, 432, 369

\bibitem[{{Semenov} {et~al.}(2005){Semenov}, {Pavlyuchenkov}, {Schreyer},
  {Henning}, {Dullemond}, \& {Bacmann}}]{Semenov05}
{Semenov}, D., {Pavlyuchenkov}, Y., {Schreyer}, K., {et~al.} 2005, \apj, 621,
  853

\bibitem[{{Semenov} \& {Wiebe}(2011)}]{Semenov11}
{Semenov}, D. \& {Wiebe}, D. 2011, \apjs, 196, 25

\bibitem[Swartz et al.(2005)]{Swartz05} Swartz, D.~A., Drake, 
J.~J., Elsner, R.~F., et al.\ 2005, \apj, 628, 811 

\bibitem[Tenenbaum et al.(2009)]{Tenenbaum09} Tenenbaum, E.~D., 
Milam, S.~N., Woolf, N.~J., \& Ziurys, L.~M.\ 2009, \apjl, 704, L108 


\bibitem[{{Teyssier} {et~al.}(2004){Teyssier}, {Foss{\'e}}, {Gerin}, {Pety},
  {Abergel}, \& {Roueff}}]{Teyssier04}
{Teyssier}, D., {Foss{\'e}}, D., {Gerin}, M., {et~al.} 2004, \aap, 417, 135

\bibitem[{{Thaddeus} {et~al.}(1985){Thaddeus}, {Vrtilek}, \&
  {Gottlieb}}]{Thaddeus85}
{Thaddeus}, P., {Vrtilek}, J.~M., \& {Gottlieb}, C.~A. 1985, \apjl, 299, L63

\bibitem[{{Thi} {et~al.}(2004){Thi}, {van Zadelhoff}, \& {van
  Dishoeck}}]{Thi04}
{Thi}, W., {van Zadelhoff}, G., \& {van Dishoeck}, E.~F. 2004, \aap, 425, 955

\bibitem[{{Walsh} {et~al.}(2012){Walsh}, {Nomura}, {Millar}, \&
  {Aikawa}}]{Walsh12}
{Walsh}, C., {Nomura}, H., {Millar}, T.~J., \& {Aikawa}, Y. 2012, \apj, 747,
  114

\bibitem[{{Willacy}(2007)}]{Willacy07}
{Willacy}, K. 2007, \apj, 660, 441

\end{thebibliography}

\clearpage

\begin{table}
\caption{$c$-C$_3$H$_2$ line results.}
\begin{center}
\begin{tabular}{l c c c c}
\hline
\hline
Transition&Frequency (GHz)&E$_{\rm u}$ (K)& Beam & $\int F dv$ (mJy km s$^{-1}$)\\
\hline
$3_{3, 0}- 2_{ 2, 1}$&216.279&19& $1\farcs3 \times 1\farcs2$(76\degree) & 53[9]\\
$6_{ 1, 6}- 5_{ 0, 5}/6_{0, 6}- 5_{1, 5}$&217.822&39& $0\farcs9 \times 0\farcs7$(83\degree) & 185[10]\\
$5_{1, 4}- 4_{2, 3}$&217.940&35& $1\farcs3 \times 1\farcs2$(78\degree) &74[9]\\ 
\hline
\end{tabular}
\end{center}
\label{tab:line}
\end{table}

\clearpage

\end{document}